# Mass Eigenstate Mixing in the 4-Generation Dirac-Kähler Extension of the SM

Alexander N. Jourjine[1]

**Abstract**

We derive the canonical form of mixing of the mass eigenstates in the lepto-quark sector of the 4-generation Dirac-Kähler extension of the SM (DK-SM) [1, 2]. The $4\times 4$ CKM matrix of DK-SM is expressed in terms two $U(2)$ matrices. It depends on 2 real parameters and 3 phases. The resulting observed $3\times 3$ CKM matrix exhibits previously unknown tree-level algebraic relations among its elements. The simplest two, $V_{ts} = V_{cb}$ and $V_{cs} = V_{tb}$, are supported by the experimental data [3]. The $4\times 4$ CKM matrix can be fully reconstructed from the experimental values of the $3\times 3$ CKM matrix. Thus, except for masses of the fourth generation, the quark sector of the DK-SM theory can be reconstructed in its entirety using the $3\times 3$ CKM matrix precision measurements.

## 1. Introduction

Dirac-Kähler (DK) spinors became interesting within the context of lattice fermion doubling problem [4]. Unlike for the Dirac spinors, their lattice naïve continuum limit turned out to coincide with the original continuum theory. They were briefly considered for description of the generation structure of the SM [5]. However, because of some technical issues with their interaction with gravity and quantization [5, 6], this approach was not pursued.

Recently, the technical problems that prevented the use of DK spinors in realistic models have been solved [1]. In [2] the mass spectrum of the DK-SM was computed and the general texture of mixing matrices was derived. In this Letter, which should be considered as a companion article to [1, 2], we address the issue of DK-SM mass eigenstate mixing.

Using the off-shell phase space representation of the action described in [2] we derive the mass mixing parameters of DK-SM in terms of spinbeins introduced in [1]. We show that the canonical DK-SM mixing matrix factorizes into product of three matrices that mix generations only pair-wise. In the quark sector the left and the right hand side factors represent $45°$ rotations of generations 1 and 4 and, separately, of the generations 2 and 3. The middle factor is a direct product of two $U(2)$ matrices that mix generations 1 and 2 and, separately, the generations 3 and 4. Mixing in the lepton sector can be represented in an even simpler way, although the simplified representation does not follow the pattern of the canonical DK-SM mixing in its left and right hand side factors. We also show that the canonical mixing agrees with the available experimental data and that the entire $4\times 4$ CKM matrix can be reconstructed from its observed $3\times 3$ CKM reduction.

The Letter is organized as follows. In Section 2 we recapitulate the results of [1, 2]. In Section 3 we derive $4\times 4$ and $3\times 3$ CKM/PMNS matrices in terms of spinbein bilinears and compare the general results with the available experimental data for $3\times 3$ CKM/PMNS matrices. We also discuss the puzzles that arise when the lepton sector is described within DK-SM. Section 4 is a summary.

---

[1] E-mail: jalnich@gmx.net. Address for correspondence: Postfach 53 01 11, 01291 Dresden, Germany.

## 2. Dirac-Kähler Extension of the SM

Given a pseudo-Riemannian manifold $M$, an on-shell massive DK spinor coupled to background gravity is defined as an inhomogeneous differential form $f$ with values in the Lie algebra of the internal symmetry group [7, 8] such that $f$ satisfies $(d - \delta - m)f = 0$, where $d$ is the exterior derivative and $\delta$ is its adjoint with respect to the scalar product of differential forms given by $\langle f, g \rangle = \int f^+ \wedge *g$ with $*$ denoting the Hodge map. Unlike for Dirac spinors, coupling of DK spinors to gravity is unique and can be defined on any manifold, including on manifolds where the spin structure cannot be defined. Given a vierbein, i. e., a set of four orthonormal frame one-forms $e^a$, and tangent space $\gamma$-matrices $\gamma^a$, define a basis, the Becher-Joos or BJ-basis, in the space of differential forms by $Z = \sum (1/p!) \gamma_{a_p} \cdots \gamma_{a_1} e^{a_1} \cdots e^{a_p}$. Then $f = Tr(\Psi Z)$ [4], where $\Psi$ is a scalar under coordinate transforms, while under local Lorentz frame rotations $e^a \to \Lambda^a_b e^b$ it transforms as $\Psi(x) \to S(\Lambda) \Psi(x) S(\Lambda)^{-1}$, where $S$ denotes a spinor representation of $SO(1,3)$. When $M$ is flat the equation on $\Psi$ that is obtained from $(d - \delta - m)f = 0$ reduces to four Dirac equations and one can identify the four columns of $\Psi$ with four generations of Dirac spinors [8]. However, such identification is possible only for flat $M$. Thus, DK spinors are physically different from Dirac spinors.

As was shown in [1], depending on the choice of a unitary spinbein gauge that defines the physical Dirac spinors, DK spinors contain different 4-multiplets of Dirac spinors. DK spinors represent a new form of fermionic matter that generalizes the notion of matter described by Dirac spinors. This is because in addition to information about the physical properties of the four Dirac spinors, a DK spinor also carries information about physical properties of the mass eigenstate mixing of the Dirac spinor constituents.

To extract Dirac degrees of freedom contained in a DK spinor in a generally covariant way one takes two sets of Dirac spinors $\xi^A$ and $\eta^A$, $A = 1,\ldots,4$, and defines the spinbein decomposition of $\Psi$ by

$$\Psi_{\alpha\beta} = \xi^A_\alpha \overline{\eta}^A_\beta, \tag{1}$$

$$\overline{\eta}^A_\alpha \eta^B_\alpha = \delta^{AB}, \quad \eta^A_\alpha \overline{\eta}^A_\beta = \delta_{\alpha\beta}, \tag{2}$$

where $\xi^A$ are taken anti-commuting and the collection of the non-dynamical commuting dimensionless normalized spinors $\eta^A$ is called the spinbein. Here $\overline{\eta}^A$ denotes the DK conjugate of $\eta^A$, $\overline{\eta}^A = \Gamma^{AB} \overline{\eta}^B$, $\overline{\eta}^B = \eta^{B+} \gamma^0$, $\Gamma \equiv diag(1, 1, -1, -1)$.

The most interesting feature of spinbein decomposition is that it allows us to assign the Dirac fields and their spinbein companions to different representations of the gauge group. This property was used in [1] to construct explicit dimension three mass terms containing Dirac spinors from different representations of the $SU(2)_L$ gauge group. Such construction is forbidden in the Standard Model but is allowed in the DK-SM.

To construct the mass terms one takes three chiral DK $SU(2)_L$ doublet fields $\Psi$, $\Phi_u$, $\Phi_d$ with spinbein decompositions that use an $SU(2)_L$ non-dynamical singlet spinbein $\eta$ and two non-dynamical factorizable $SU(2)_L$ doublet spinbeins $\chi$ and $\theta$

$$\Psi = Q\overline{\overline{\eta}}, \qquad \Phi_u = u\overline{\overline{\chi}}, \qquad \Phi_d = d\overline{\overline{\theta}}, \tag{3}$$

$$\chi_i = \varphi_i \hat{\chi}, \qquad \overline{\overline{\hat{\chi}}}\hat{\chi} = 1, \qquad \theta_i = \varphi_i^* \hat{\theta}, \qquad \overline{\overline{\hat{\theta}}}\hat{\theta} = 1, \qquad \varphi_i^* \varphi_i = 1, \tag{4}$$

where $\varphi_i$ is a $SU(2)_L$ doublet of scalars.

The left-handed $Q^A$, $A = 1,\ldots,4$, and the right-handed $u^A, d^A$ quark fields are assigned to the same symmetry representations as in the SM. Additionally, we take $\eta$ as $SU(2)_L$ singlet, but $\chi, \theta$ are $SU(2)_L$ doublets with gauge transforms

$$Q \to TQ, \; \chi \to T\chi, \; \theta \to T^*\theta, \qquad T, T^* \in SU(2), \tag{5}$$

so that under $SU(2)_L$ the three DK fields transform as doublets

$$\Psi \to T\Psi, \Phi_u \to T^*\Phi_u, \Phi_d \to T\Phi_d. \tag{6}$$

The quark sector of the DK-SM is then given by the $SU(2)_L$ invariant Lagrangian

$$\mathcal{L} = Tr\left[\overline{\overline{\Psi}}(i\rlap{D}\mkern1mu/\,)\Psi + \overline{\overline{\Phi}}_u(i\rlap{D}\mkern1mu/\,)\Phi_u + \overline{\overline{\Phi}}_d(i\rlap{D}\mkern1mu/\,)\Phi_d\right] - \\ - m_u Tr\left[\overline{\overline{\Psi}}(i\sigma^2)\Phi_u - \overline{\overline{\Phi}}_u(i\sigma^2)\Psi\right] - m_d Tr\left[\overline{\overline{\Psi}}\Phi_d + \overline{\overline{\Phi}}_d\Psi\right], \tag{7}$$

where $\rlap{D}\mkern1mu/\,$ are the appropriate gauged derivatives and parameters $m_u, m_d$ are bare masses for the up and down quarks [1]. Note the absence of Yukawa couplings with the Higgs field. Fermion mass generation in DK-SM does not require Higgs field. In the unitary gauge [1] with

$$\eta = const, \; \hat{\chi} = const, \; \hat{\theta} = const \tag{8}$$

we obtain

$$\mathcal{L} = \mathcal{L}_\mathcal{K} + \mathcal{L}_\mathcal{M},$$
$$\mathcal{L}_\mathcal{K} = \overline{\overline{Q}}_i^A(i\rlap{D}\mkern1mu/\,)Q_i^A + \overline{\overline{u}}^A(i\rlap{D}\mkern1mu/\,)u^A + \overline{\overline{d}}^A(i\rlap{D}\mkern1mu/\,)d^A, \tag{9}$$
$$\mathcal{L}_\mathcal{M} = -m_u\left(\mathcal{M}_u^{AB}\overline{\overline{Q}}_i^A(i\sigma^2)\varphi_i^* u^B - \tilde{\mathcal{M}}_u^{AB}\overline{\overline{u}}^B(i\sigma^2)\varphi_i Q_i^A\right) - m_d\left(\mathcal{M}_d^{AB}\overline{\overline{Q}}_i^A \varphi_i d^B + \tilde{\mathcal{M}}_d^{AB}\overline{\overline{d}}^A\varphi_i^* Q_i^B\right),$$
$$\mathcal{M}_u^{AB} = \overline{\overline{\hat{\chi}}}^B \eta^A, \qquad \tilde{\mathcal{M}}_u^{AB} = \overline{\overline{\eta}}^B \hat{\chi}^A, \qquad \mathcal{M}_d^{AB} = \overline{\overline{\hat{\theta}}}^B \eta^A, \qquad \tilde{\mathcal{M}}_d^{AB} = \overline{\overline{\eta}}^B \hat{\theta}^A.$$

We observe that non-dynamical $\varphi$ plays the role of the classical vacuum field of the Higgs field of the SM. We also note that using the unitary spinbein gauge (8) generates mass matrices $m_u \mathcal{M}_u^{AB}$, $m_d \mathcal{M}_d^{AB}$ that are exactly the same as those obtained in the SM using Yukawa terms for quarks. When $\varphi_1 = 0$, $\varphi_2 = 1$ we obtain

$$\mathcal{L} = \overline{Q}_i^A (i\slashed{D}) Q_i^A + \overline{u}^A (i\slashed{D}) u^A + \overline{d}^A (i\slashed{D}) d^A - \left( m_u \overline{Q}_1^A \mathcal{M}_u^{AB} u^B + m_d \overline{Q}_2^A \mathcal{M}_d^{AB} d^B + c.c. \right). \quad (10)$$

The spectrum of the DK-SM theory was computed in [2]. The computation uses the Cartan group element decomposition for $\mathcal{M} \in U(2,2)$ that is given by $\mathcal{M} = U R V$, where $U, V \in U(2) \times U(2)$ and

$$R = \begin{pmatrix} C & S \\ S & C \end{pmatrix}, \quad S = diag(sh\lambda_1, sh\lambda_2), \quad C = diag(ch\lambda_1, ch\lambda_2), \quad (11)$$

with $\lambda_1, \lambda_2$ $\lambda_1 \geq \lambda_2$, non-negative real. The eigenvalues of $R$ are $(e^{-\lambda_1}, e^{-\lambda_2}, e^{+\lambda_1}, e^{+\lambda_2})$. The lepto-quark mass spectrum of the DK-SM is given by eight masses

$$\begin{aligned} \{m_{u,d}^A\} &= (m_{u,d} e^{-\lambda_1}, m_{u,d} e^{-\lambda_2}, m_{u,d} e^{+\lambda_1}, m_{u,d} e^{+\lambda_2}), \\ \{m_{\nu,e}^A\} &= (m_{\nu,e} e^{-\lambda_3}, m_{\nu,e} e^{-\lambda_4}, m_{\nu,e} e^{+\lambda_3}, m_{\nu,e} e^{+\lambda_4}), \end{aligned} \quad (12)$$

that trivially satisfy four mass relations

$$m_u^1 m_u^3 = m_u^2 m_u^4, \quad m_d^1 m_d^3 = m_d^2 m_d^4, \quad m_\nu^1 m_\nu^3 = m_\nu^2 m_\nu^4, \quad m_e^1 m_e^3 = m_e^2 m_e^4, \quad (13)$$

where $m_{u,d}$, $m_{\nu,e}$ are the lepto-quark bare masses. Note that this particular ordering of the eigenvalues of $R$ implies that $m_{u,d}^3 \geq m_{u,d}^4$. Below, when we shall discuss the experimental evidence for DK-SM, we shall change the ordering to the conventional one. Comparison with the experimental mass values from [3] results in the DK-SM mass related parameters given in Fig. 1 in the conventionally ordered form. Note that masses of 4$^{th}$ quarks are given only within order of magnitude accuracy.

To compute mass eigenstate mixing parameters we use the off-shell Fourier expansions for the DK fields [2]

$$\begin{aligned} \xi_L^A(x) &= \int \frac{d^4 k}{(2\pi)^4} \left( b_2^A(k) w_{L-}(k) e^{-ikx} + d_1^{A+}(k) w_{L+}(k) e^{ikx} \right), \\ \xi_R^A(x) &= \int \frac{d^4 k}{(2\pi)^4} \left( b_1^A(k) w_{R-}(k) e^{-ikx} + d_2^{A+}(k) w_{R+}(k) e^{ikx} \right), \quad A = 1, 2, \\ \xi_L^A(x) &= \int \frac{d^4 k}{(2\pi)^4} \left( b_2^{A+}(k) w_{L-}(k) e^{-ikx} + d_1^A(k) w_{L+}(k) e^{ikx} \right), \\ \xi_R^A(x) &= \int \frac{d^4 k}{(2\pi)^4} \left( b_1^{A+}(k) w_{R-}(k) e^{-ikx} + d_2^A(k) w_{R+}(k) e^{ikx} \right), \quad A = 3, 4, \end{aligned} \quad (14)$$

where $(\xi_L^A(x), \xi_R^A(x))$ stand for $(Q_1^A(x), u^A(x))$ or $(Q_2^A(x), d^A(x))$ in (10). The four off-shell Dirac spinor amplitudes $w_{L,R\pm}(k)$ are given by

$$\begin{aligned} w_{L-}(k) &= (k \cdot \sigma)^{1/2} u, & w_{L+}(k) &= (k \cdot \sigma)^{1/2^+} v, \\ w_{R-}(k) &= (k \cdot \bar{\sigma})^{1/2^+} v, & w_{R+}(k) &= -(k \cdot \bar{\sigma})^{1/2} u, \end{aligned} \quad (15)$$

where we use the chiral representation of the $\gamma$-matrices with $\gamma^5 = -\Gamma$ [9], $u, v$ are two constant orthonormal spinors and $k \cdot \sigma = k_\mu \sigma^\mu$, $k \cdot \bar{\sigma} = k_\mu \bar{\sigma}^\mu$, $\sigma^\mu = (1, \sigma^i)$, $\bar{\sigma}^\mu = (1, -\sigma^i)$. Here $\sigma^i$, $i = 1, 2, 3$, are the Pauli matrices. In the reference frame with $k_1 = k_2 = 0$ we can take for up spin $u = (0,1)^T$ and for down spin $v = (1,0)^T$. The square root of Hermitean matrices $(k \cdot \sigma)$, $(k \cdot \bar{\sigma})$ is uniquely defined via positive sign square roots of their eigenvalues. On-shell (15) reduces to the standard expressions [9]. The off-shell amplitudes satisfy

$$w_{L-}^+(k)(k \cdot \bar{\sigma}) w_{L+}(k) = 0, \qquad w_{R-}^+(k)(k \cdot \sigma) w_{R+}(k) = 0, \qquad (16)$$
$$w_{L-}^+(k) w_{R-}(k) = 0, \qquad w_{L+}^+(k) w_{R+}(k) = 0.$$

|         | $\nu$                  | $e$              | $d$                   | $u$                    |
|---------|------------------------|------------------|-----------------------|------------------------|
| $\lambda_1$ | $> 29$              | 9.0              | $4.6 - 5.3$           | $8.3 - 9.1$            |
| $\lambda_2$ | $0.82 - 0.93$       | 1.4              | $1.5 - 2.3$           | $1.7 - 2.5$            |
| $m$     | $(2.0 - 2.1)\cdot 10^{-11}$ | 0.43      | $0.57 - 0.79$         | $14 - 15$              |
| $m_1$   | $< 10^{-24}$           | $5.1 \cdot 10^{-4}$ | $(4.1 - 5.8) \cdot 10^{-3}$ | $(1.7 - 3.3) \cdot 10^{-3}$ |
| $m_2$   | $(8.6 - 8.8) \cdot 10^{-12}$ | 0.11      | $0.08 - 0.13$         | 1.3                    |
| $m_3$   | $(48 - 50) \cdot 10^{-12}$ | 1.8          | $(4.1 - 4.8)$         | 170                    |
| $m_4$   | $> 100$                | 370              | $(60 - 150)$          | $(60 - 140) \cdot 10^3$ |

Fig. 1. Masses (GeV) and related parameters in the DK-SM. The underlined mass values are DK-SM estimates. The remaining mass values, quoted with two digit precision, are from [3]. The spread for the strange quark mass is the combined spread for MS and 1S subtraction schemes.

## 3. Mass Eigenstate Mixing in the DK-SM

We now proceed with deriving the $4 \times 4$ CKM matrix. First we define the anti-symmetrized normal ordering of Grassmann bilinear forms (a-normal ordering), denoted as $\vdots \ \vdots$, as the difference between the normal ordering $:\ :$ and $:\ :_A$, the anti-normal ordering, where all creation and annihilation operators switch their positions compared to normal ordering, that is $\vdots \ \vdots \equiv 1/2(:\ : - :\ :_A)$.

Then in the phase space the up quark component of the action (10) can be written as a bilinear form in the 32-dimensional Grassmann space according to

$$\vdots S \vdots = S_0 + S_i,$$
$$S_0 = \frac{1}{2}\int \frac{d^4k}{(2\pi)^4}\, C^+(k)\begin{pmatrix} \mathcal{S}_1(k) & 0 \\ 0 & \mathcal{S}_2(k) \end{pmatrix} C(k), \qquad (17)$$
$$S_i = \frac{1}{2}\int \frac{d^4k}{(2\pi)^4}\frac{d^4k'}{(2\pi)^4}\, C^+(k)\begin{pmatrix} \mathcal{I}_{11}(k,k') & \mathcal{I}_{12}(k,k') \\ \mathcal{I}_{21}(k,k') & \mathcal{I}_{22}(k,k') \end{pmatrix} C(k'),$$

where $S_0$, $S_i$ are the free and the interaction parts, respectively, and $C(k) = (C_1(k),\ C_2(k))$. The phase space down quark component of the action (10) has the same form. The exact form of the interaction matrix $\mathcal{I}_{ik}(k,k')$, $i,k=1,2$ is not needed for our derivations. The two sixteen-component vectors $C_\alpha$ with definite spin $\alpha = 1,2$ are defined by

$$C_\alpha(k) = \left(b_\alpha^A(k), d_\alpha^A(-k), b_\alpha^{A+}(k), d_\alpha^{A+}(-k)\right)^T. \qquad (18)$$

Since below we will be dealing with manifestly covariant expressions for various quantities, for simplicity from now on we shall use the $k_1 = k_2 = 0$ reference frame. The two densities $\mathcal{S}_\alpha(k) = \mathcal{S}_\alpha(k, M)$, $\alpha = 1,2$, where $M = f\, R_d$, satisfy $\mathcal{S}_\alpha^+(k, M) = \mathcal{S}_\alpha(k, M)$ and $\mathcal{S}_2(k, M) = \mathcal{S}_1(k, M^+)$. The use of the Fourier expansion (14) results in

$$\mathcal{S}_1(k) = \begin{pmatrix} g_-\Gamma & M_c^+ & 0 & M_d^+ \\ M_c & g_+\Gamma & -M_d^+ & 0 \\ 0 & -M_d & -g_-\Gamma & M_c^* \\ M_d & 0 & M_c^T & -g_+\Gamma \end{pmatrix}, \qquad \mathcal{S}_2(k) = (\mathcal{S}_1(k))^*, \qquad (19)$$

where $\Gamma = diag(1,\ 1,-1,-1)$ is the matrix that defines $U(2,2)$, factors $g_\pm(k)$, $f(k)$ are

$$g_\pm(k) = \mp w_{L\pm}^+(k)(k\cdot\bar\sigma)w_{L\pm}(k) = \mp w_{R\pm}^+(k)(k\cdot\sigma)w_{R\pm}(k) = \mp sign(k_0 \mp k_3)|k^2|,$$
$$f(k) = w_{L+}^+(-k)\, w_{R-}(k) = w_{L-}^+(k)\, w_{R+}(-k) = i(k_0 - k_3)^{1/2}(k_0 + k_3)^{1/2*}, \qquad (20)$$

and matrices $M_d$, $M_c$ are given by

$$M_d = \begin{pmatrix} -fC & 0 \\ 0 & f^*C \end{pmatrix}, \qquad C = \begin{pmatrix} m\,ch\lambda_1 & 0 \\ 0 & m\,ch\lambda_2 \end{pmatrix}, \tag{21}$$

$$M_c = \begin{pmatrix} 0 & f^*S \\ -fS & 0 \end{pmatrix}, \qquad S = \begin{pmatrix} m\,sh\lambda_1 & 0 \\ 0 & m\,sh\lambda_2 \end{pmatrix}. \tag{22}$$

Here we carried out the $U(2) \times U(2)$ flavor rotations on the quark fields $Q_i^A, u^A, d^A$ so that mass matrices $\mathcal{M}_{u,d}$ in (10) have been reduced to their middle factors (11).

It turns out that in $S_0$ in (17) the elements $(1),(4),(6),(7)$ of the $C_1(k)$ vector form one 4-dimensional invariant subspace, while elements $(2),(3),(5),(8)$ form an orthogonal 4-dimensional invariant subspace. This allows us to write $16 \times 16$ matrices $\mathcal{S}_\alpha(k)$ in $4 \times 4$ block-diagonal form

$$\mathcal{S}_\alpha(k) = diag\left(\mathcal{S}_\alpha^{(+)}(k,\lambda_1), \mathcal{S}_\alpha^{(-)}(k,\lambda_1), \mathcal{S}_\alpha^{(+)}(k,\lambda_2), \mathcal{S}_\alpha^{(-)}(k,\lambda_2)\right), \tag{23}$$

where $\mathcal{S}_1^{(\pm)}(k,\lambda_k)$ are given by

$$\mathcal{S}_1^{(+)}(k,\lambda) = \begin{pmatrix} g_- & -f^*s & 0 & -f^*c \\ -fs & -g_+ & -fc & 0 \\ 0 & -f^*c & g_- & -f^*s \\ -fc & 0 & -fs & -g_+ \end{pmatrix}, \quad \mathcal{S}_1^{(-)}(k,\lambda) = -\mathcal{S}_1^{(+)}(f^*,k,\lambda), \tag{24}$$

where now $c = m\,ch\lambda, s = m\,sh\lambda$, $\lambda = \lambda_{1,2}$ and similar for $\mathcal{S}_2^{(\pm)}(k)$. Additional renaming of rows and columns according to $1 \leftrightarrow 1, 2 \leftrightarrow 3, 4 \leftrightarrow 4$ reduces $\mathcal{S}_\alpha^{(\pm)}(k)$, $\alpha = 1,2$, to the desired $2 \times 2$ block form. We obtain that $\mathcal{S}_1^{(+)}(k,\lambda)$ given by

$$\mathcal{S}_1^{(+)}(k,\lambda) = \begin{pmatrix} g_- & 0 & -f^*s & -f^*c \\ 0 & g_- & -f^*c & -f^*s \\ -fs & -fc & -g_+ & 0 \\ -fc & -fs & 0 & -g_+ \end{pmatrix}. \tag{25}$$

We now want to reduce (25) to the form generated by two non-interacting Dirac spinors with no mixing. This can be done using a similarity transformation with the $4 \times 4$ unitary matrix $W_B$ given by

$$W_B = \begin{pmatrix} V_2 & 0 \\ 0 & V_2 \end{pmatrix}, \qquad V_2 = \frac{1}{\sqrt{2}} \begin{pmatrix} 1 & 1 \\ -1 & 1 \end{pmatrix}. \tag{26}$$

It generates the desired representation of the phase space action that up to the sign of the masses is identical to that of two non-mixing DK spinors with masses $m\,e^{\mp\lambda}$, $m = m_{u,d}$, $\lambda = \lambda_{1,2}$. We obtain that $\mathcal{S}_1^{(+)}(k,\lambda)$ becomes

$$\widetilde{\mathcal{S}}_1^{(+)}(k,\lambda) = \begin{pmatrix} g_- & 0 & -f^*m e^{-\lambda} & 0 \\ 0 & g_- & 0 & f^*m e^{\lambda} \\ -f m e^{-\lambda} & 0 & -g_+ & 0 \\ 0 & f m e^{\lambda} & 0 & -g_+ \end{pmatrix}. \tag{27}$$

We now proceed with deriving the DK-SM mass eigenstate mixing matrix. First, we change the ordering of masses from the original $\left(e^{-\lambda_1}, e^{-\lambda_2}, e^{+\lambda_1}, e^{+\lambda_2}\right)$ to the one used in the SM in the order of increasing mass to $\left(e^{-\lambda_1}, e^{-\lambda_2}, e^{+\lambda_2}, e^{+\lambda_1}\right)$. In practice this means switching the 3$^d$ and the 4$^{th}$ generation indices. As a result matrix $V_B$ in (26) now takes the form

$$W_B = \frac{1}{\sqrt{2}} \begin{pmatrix} 1 & 0 & 0 & 1 \\ 0 & 1 & 1 & 0 \\ 0 & -1 & 1 & 0 \\ -1 & 0 & 0 & 1 \end{pmatrix}. \tag{28}$$

Before we proceed with deriving the canonical form of the DK-SM mixing a few remarks about our choice of the ordering for the off-shell action (17) are in order. Firstly, we are forced to use the a-normal ordering, because only then we can carry out the unitary Bogoliubov transformations that mix the creation and annihilation operators. Secondly, the flipped anti-commutation relations for the last two generations and the choice of ordering explain the appearance of the unitary group $U(2) \times U(2)$ in the action that, when commuting fields are used, has an entirely different $U(2,2)$ symmetry group. It only with our choice of ordering, that we are able to diagonalize the quark mass matrices using unitary transformations. Classically this is not possible. Classically, mass matrices must be diagonalized using the $U(2,2)$ symmetry and the result is the complete mass degeneracy of classical DK-SM [2]. Therefore, the resulting $(U(2) \oplus U(2)) \times (U(2) \oplus U(2))$ symmetry of the massless DK action appears only on quantum level and only when the a-normal ordering is used.

Turning to mixing, the quark mixing matrix is defined [3] as

$$W^{DK} = W_L^u W_L^{d+}. \tag{29}$$

$W_L^{u,d}$ contains two factors. The factor on the right comes from $U(2) \times U(2)$ of premixing of the first two generations and, separately, of the last two generations. The factor on the left comes from Bogoliubov mixing (28). Therefore,

$$W_L^{u,d} = W_{14,23}^{u,d} W_{12,34}^{u,d}, \tag{30}$$

where index $ij, kl$ denotes $U(2)$ mixing of generations $ij$ and, separately, generations $kl$, we obtain that the full mixing matrix has the form

$$W^{DK} = W_{14,23}^u \widetilde{W}_{12,34}^{ud} \left(W_{14,23}^d\right)^+, \qquad W_B^u \equiv W_{14,23}^u, \qquad W_B^d \equiv W_{14,23}^d,$$
$$\widetilde{W}_{12,34}^{ud} = W_{12,34}^u \left(W_{12,34}^d\right)^+ \equiv W_P \in U(2) \oplus U(2) \subset U(4), \tag{31}$$

where $W_B^{u,d}$, $W_P$ belong to two different $U(2)\oplus U(2)$ subgroups of flavor $U(4)$. Similar decomposition exists for the leptonic sector. We shall call $W_B^{u,d}$ the Bogoliubov or b-mixing matrices. The mix only $A=1,4$ and separately $A=2,3$ generations and are given by the same matrix (28)

$$W_B^u = W_B^d = W_B = \frac{1}{\sqrt{2}}\begin{pmatrix} 1 & 0 & 0 & 1 \\ 0 & 1 & 1 & 0 \\ 0 & -1 & 1 & 0 \\ -1 & 0 & 0 & 1 \end{pmatrix}. \tag{32}$$

At the same time $W_P$, that we shall call the premixing or p-mixing matrix, mixes only $A=1,2$ and separately $A=3,4$ generations and has the form

$$W_P = \begin{pmatrix} x_1 & y_1 & 0 & 0 \\ z_1 & w_1 & 0 & 0 \\ 0 & 0 & x_2 & y_2 \\ 0 & 0 & z_2 & w_2 \end{pmatrix}, \quad W_P \in U(2)\oplus U(2) \subset U(4). \tag{33}$$

Using (29-33) we obtain the canonical form of the CKM mixing matrix in the DK-SM theory.

$$W^{DK} = W_B W_P (W_B)^T. \tag{34}$$

It is important to note that the order of the factors in (30) is fixed. This is because we performed the removal of the $U(2)\times U(2)$ factors in $\mathcal{M}_{u,d}^{AB}$ first and only then applied the Bogoliubov transformation to diagonalize $U(2,2)$ mass matrices.

Substitution of (32, 33) into (34) results in the DK-SM $4\times 4$ CKM mass eigenstate mixing matrix (CKM-4) that is given by

$$W^{DK} = \frac{1}{2}\begin{pmatrix} x_1+w_2 & y_1+z_2 & -y_1+z_2 & -x_1+w_2 \\ z_1+y_2 & w_1+x_2 & -w_1+x_2 & -z_1+y_2 \\ -z_1+y_2 & -w_1+x_2 & w_1+x_2 & z_1+y_2 \\ -x_1+w_2 & -y_1+z_2 & y_1+z_2 & x_1+w_2 \end{pmatrix}. \tag{35}$$

Taking the upper $3\times 3$ block of (35) we obtain the currently visible $3\times 3$ non-unitary mixing matrix CKM-3

$$W_{SM}^{DK} = \frac{1}{2}\begin{pmatrix} x_1+w_2 & y_1+z_2 & -y_1+z_2 \\ z_1+y_2 & w_1+x_2 & -w_1+x_2 \\ -z_1+y_2 & -w_1+x_2 & w_1+x_2 \end{pmatrix}. \tag{36}$$

We now turn to the question of the experimental evidence for mass eigenstate mixing described by (35, 36). We begin with the quark sector. First, we shall determine the number of independent parameters of $W_{SM}^{DK}$ that remain after removal of its phases by rephrasing the quark fields. Obviously, the results will also apply to the lepton sector with Dirac neutrinos.

Since Bogoliubov factors are constant matrices, the number of parameters in $W_{SM}^{DK}$ is determined by the two $U(2)$ factors. There are eight of them: two real parameters and six phases. This can be seen from the explicit expression for a $U(2)$ matrix

$$U_2 = \begin{pmatrix} x & y \\ z & w \end{pmatrix} = \begin{pmatrix} \cos\chi\, e^{i\alpha} & -\sin\chi\, e^{i\beta} \\ \sin\chi\, e^{i\gamma} & \cos\chi\, e^{i\delta} \end{pmatrix}, \quad \delta = -\alpha + \beta + \gamma. \tag{37}$$

After extraction of the phases in two $U(2)$ factors of $W_P$ and application of the available 7 phases of rephrasing of the quark fields the CKM-4 can be reduced to

$$W^{DK} = \frac{1}{2} \begin{pmatrix} 1 & 0 & 0 & e^{i\Lambda_1} \\ 0 & 1 & e^{i\Lambda_2} & 0 \\ 0 & -e^{-i\Lambda_2} & 1 & 0 \\ -e^{-i\Lambda_1} & 0 & 0 & 1 \end{pmatrix} \begin{pmatrix} c_1 & -s_1 & 0 & 0 \\ s_1 & c_1 & 0 & 0 \\ 0 & 0 & c_2 & -s_2 \\ 0 & 0 & s_2 & c_2 \end{pmatrix} \begin{pmatrix} 1 & 0 & 0 & -e^{-i\bar{\Lambda}_1} \\ 0 & 1 & -e^{-i\bar{\Lambda}_2} & 0 \\ 0 & e^{i\bar{\Lambda}_2} & 1 & 0 \\ e^{i\bar{\Lambda}_1} & 0 & 0 & 1 \end{pmatrix}, \tag{38}$$

where $c_1 \equiv \cos\chi_1$, *etc*, and

$$\begin{aligned} \Delta_1 &= \beta_1 + \alpha_2 - \beta_2 - \gamma_2, & \bar{\Delta}_1 &= \alpha_1 - \beta_1, \\ \Delta_2 &= -\alpha_1 + \beta_1 + \gamma_1 - \beta_2, & \bar{\Delta}_2 &= -\alpha_2 + \beta_2. \end{aligned} \tag{39}$$

It follows from (38) that $W^{DK}$ contains two real parameters $\chi_k, k=1,2$ and three phases derived from $\Delta_k, \bar{\Delta}_k$ after extracting common $U(1)$ factors. Indeed, for our purposes we can take each factor in (34) belongs to $SU(2)$. Then in (37) $\beta = -\gamma$ and we obtain a constraint on $\Delta_k, \bar{\Delta}_k$ that defines the three independent phases

$$\Delta_1 + \Delta_2 + \bar{\Delta}_1 + \bar{\Delta}_2 = 0. \tag{40}$$

We now can analyze the experimental evidence for $V_{CKM}^{DK}$ (CKM-3 matrix) defined in (36)

$$V_{CKM}^{DK} = \frac{1}{2} \begin{pmatrix} x_1 + w_2 & y_1 + z_2 & -y_1 + z_2 \\ z_1 + y_2 & w_1 + x_2 & -w_1 + x_2 \\ -z_1 + y_2 & -w_1 + x_2 & w_1 + x_2 \end{pmatrix}. \tag{36}$$

The CKM$_4$ matrix $W^{DK}$ and, therefore, CKM$_3$ matrix $V_{CKM}^{DK}$ depend on two real and three phase parameters. This should be compared to three real plus one phase parameters of the SM and to six real and three phase parameters of the CKM-4 in SM-4. Therefore, $V_{CKM}^{DK}$ is more economical then the general SM CKM matrix in real parameters but offers two more phases. This can be beneficial for accounting for CP violation needed to explain matter-antimatter asymmetry. As is well-known, the single phase of the SM CKM does not generate the sufficient effect. Clearly, $V_{CKM}^{DK}$ is far more economical then CKM-4 of SM-4.

The second observation is that using (34) we can explain the smallness of the mixing between the 3d and the first two generations by assuming that for quark sector

$$U_2^1 = \begin{pmatrix} x_1 & y_1 \\ z_1 & w_1 \end{pmatrix} \approx \begin{pmatrix} x_2 & y_2 \\ z_2 & w_2 \end{pmatrix} = U_2^2. \tag{41}$$

Then, because the lower-left and upper-right blocks involve subtraction of the elements $U_2^1$ and $U_2^2$ in (41), the CKM$_4$ matrix elements in these blocks will be suppressed. In (36) the relation (41) leads to suppression of $V_{td}, V_{ts}, V_{ub}, V_{cb}$ compared to the remaining elements. The suppression is indeed observed experimentally [3]. As we shall see below the lepton sector exhibits a remarkably different pattern that leads to the near-tri-bimaximal mixing pattern.

Further, the CKM-3 mixing matrix $V_{CKM}^{DK}$ in (36) exhibits a number of algebraic constraints imposed on its elements by its form. (Of course, such constraints also exist in Sm CKM-3. However, they have different functional form.) The simplest two are

$$V_{ts} = V_{cb}, \tag{42}$$
$$V_{cs} = V_{tb}. \tag{43}$$

The derived relation $|V_{ts}| = |V_{cb}|$ agrees within the experimental data in two different ways. The first agreeement comes from separate measurement of both sides of (42). We quote

$$\begin{aligned}|V_{ts}| &= (38.7 \pm 2.1) \cdot 10^{-3} \quad [3], \\ |V_{cb}| &= (38.7 \pm 1.1) \cdot 10^{-3} \quad [10], \qquad |V_{cb}| = (41.5 \pm 0.7) \cdot 10^{-3} \quad [10].\end{aligned} \tag{44}$$

The second agreeement comes from the measurement of the ratio of the elements

$$|V_{ts}/V_{cb}| = 1.04 \pm 005 \quad [3]. \tag{45}$$

Note, that derivation of $|V_{ts}|$ involves loop calculations and thus explicitly uses the SM CKM-3 unitarity. Thus our claim of partial experimental verification of (42) can only hold if the inevitable violation of $3 \times 3$ unitarity through the fourth generation is sufficiently small. That it is relatively small follows from (41).

The constraint (43) is also in agreement with the data, although not with comparable precision. Direct measurement of $V_{cs}$ from two different experiments gives [3]

$$\begin{aligned}(a) \quad V_{cs} &= 1.030 \pm 0.038, \\ (b) \quad V_{cs} &= 0.98 \pm 0.01 \pm 0.10,\end{aligned} \tag{46}$$

where for (b) the first/second error is experimental/theoretical, while direct measurement of $V_{tb}$ gives [3]

$$V_{tb} = 0.88 \pm 0.07. \tag{47}$$

Note that the constraints (42, 43) are given at tree level. Because of radiative corrections, they are not expected to hold exactly.

The remaining algebraic constrains on the CKM$_3$ matrix can be derived from the measured values of the quark mixing matrix as follows. First we note that the elements of $U_2^2$ in (41) can be expressed in terms of elements of $U_2^1$ and the $V_{31}, V_{32}, V_{13}$ elements of the measured CKM$_3$ matrix (denoted below as $V$) according to

$$x_2 = 2V_{32} + w_1, \qquad y_2 = 2V_{31} + z_1,$$

$$z_2 = 2V_{13} + y_1, \qquad w_2 = -(2V_{13} + y_1)\frac{2V_{32}^* + w_1^*}{2V_{31}^* + z_1^*}, \tag{48}$$

where $w_2$ was obtained using the unitarity of $U_2^2$. Now we can determine the elements of $U_2^1$ using the $V_{11}, V_{12}, V_{21}, V_{22}$ elements of $V$ to obtain

$$x_1 = 2V_{11} + V_{12}\frac{V_{22}^*}{V_{21}^*} \approx V_{11}, \qquad y_1 = V_{12} - V_{13} \approx V_{12},$$

$$z_1 = V_{21} - V_{31} \approx V_{21}, \qquad w_1 = V_{22} - V_{32} \approx V_{22}, \tag{49}$$

where we used the experimental values for $V$ to write down the approximate values on the right hand side of the equations. System (49) implies that the upper left block of CKM$_3$ matrix is approximately $2 \times 2$ unitary, which is indeed observed in the data [3].

Using the unitarity of $U_2^1$, from (49) we obtain four real constraints on $V$

$$|x_1|^2 + |y_1|^2 = 1,$$

$$x_1 z_1^* + y_1 w_1^* = 0, \tag{50}$$

$$|z_1|^2 + |w_1|^2 = 1.$$

Leaving a detailed analysis of these constraints to another publication, we shall consider only the simplest constraint $|z_1|^2 + |w_1|^2 = 1$. From (49, 50) we obtain

$$\Delta V \equiv |V_{21} - V_{31}|^2 + |V_{22} - V_{32}|^2 = 1. \tag{51}$$

Since the values of the phases of the elements of $V$ in (49) are known with insufficient precision we shall only compute the minimal and the maximal values of the left hand side of (51). We obtain, using the known values of the absolute values of $V$ from [3], that

$$\Delta V_{min} = 0.9347,$$
$$\Delta V_{max} = 1.0941. \tag{52}$$

The values indicate that the phase differences in (51) find themselves in the acceptable region $0° - 180°$ with their likely values in the middle of the region. Note that the constraint (51) is non-trivial. It involves only the elements from the lower-left $2 \times 2$ block of $V$ and cannot be reduced to a unitarity condition.

We now turn to the lepton sector mixing. The tri-bimaximal PMNS mixing matrix is given by [11]

$$U_{TBM} = \begin{pmatrix} \sqrt{2/3} & \sqrt{1/3} & 0 \\ -\sqrt{1/6} & \sqrt{1/3} & -\sqrt{1/2} \\ \sqrt{1/6} & -\sqrt{1/3} & -\sqrt{1/2} \end{pmatrix}, \tag{53}$$

where we used the freedom to redefine phases of lepton fields to multiply the third row with $-1$.

The experimental data [12] shows that the observed lepton mixing pattern is near-tri-bimaximal. It is easy to see that we can represent the PMNS-4, the $4 \times 4$ DK-SM lepton mixing matrix, in the factorized form as

$$U_{PMNS}^{DK} = \begin{pmatrix} 1 & 0 & 0 & 0 \\ 0 & 1/\sqrt{2} & 1/\sqrt{2} & 0 \\ 0 & -1/\sqrt{2} & 1/\sqrt{2} & 0 \\ 0 & 0 & 0 & 1 \end{pmatrix} \begin{pmatrix} x_1 & y_1 & 0 & 0 \\ z_1 & w_1 & 0 & 0 \\ 0 & 0 & x_2 & y_2 \\ 0 & 0 & z_2 & w_2 \end{pmatrix} \begin{pmatrix} 1 & 0 & 0 & 0 \\ 0 & 1 & 0 & 0 \\ 0 & 0 & 1 & 0 \\ 0 & 0 & 0 & 1 \end{pmatrix}. \tag{54}$$

From (54) we obtain PMNS-4 mixing matrix

$$U_{PMNS}^{DK} = \begin{pmatrix} x_1 & y_1 & 0 & 0 \\ z_1/\sqrt{2} & w_1/\sqrt{2} & x_2/\sqrt{2} & y_2/\sqrt{2} \\ -z_1/\sqrt{2} & -w_1/\sqrt{2} & x_2/\sqrt{2} & y_2/\sqrt{2} \\ 0 & 0 & z_2 & w_2 \end{pmatrix}, \tag{55}$$

and the PMNS-3 mixing matrix

$$U_{PMNS}^{DK} = \begin{pmatrix} x_1 & y_1 & 0 \\ z_1/\sqrt{2} & w_1/\sqrt{2} & x_2/\sqrt{2} \\ -z_1/\sqrt{2} & -w_1/\sqrt{2} & x_2/\sqrt{2} \end{pmatrix}. \tag{56}$$

Comparing (56) and (53) we obtain that the tri-bimaximal pattern of (53) is reproduced if

$$x_1 = \sqrt{2/3}, \quad y_1 = \sqrt{1/3}, \quad x_2 = -1, \quad y_2 = 0,$$
$$z_1 = -\sqrt{1/3}, \quad w_1 = \sqrt{2/3}, \quad z_2 = 0, \quad w_2 = \pm 1. \tag{57}$$

We observe that the lepton sector with Dirac neutrinos offers a remarkably different mixing pattern. At tree level for quarks we obtained

$$V_{CKM}^{DK} \approx \begin{pmatrix} 1/\sqrt{2} & 0 & 0 & 1/\sqrt{2} \\ 0 & 1/\sqrt{2} & 1/\sqrt{2} & 0 \\ 0 & -1/\sqrt{2} & 1/\sqrt{2} & 0 \\ -1/\sqrt{2} & 0 & 0 & 1/\sqrt{2} \end{pmatrix} \begin{pmatrix} x & y & 0 & 0 \\ z & w & 0 & 0 \\ 0 & 0 & x+\varepsilon_x & y+\varepsilon_y \\ 0 & 0 & z+\varepsilon_z & w+\varepsilon_w \end{pmatrix} \begin{pmatrix} 1/\sqrt{2} & 0 & 0 & -1/\sqrt{2} \\ 0 & 1/\sqrt{2} & -1/\sqrt{2} & 0 \\ 0 & 1/\sqrt{2} & 1/\sqrt{2} & 0 \\ 1/\sqrt{2} & 0 & 0 & 1/\sqrt{2} \end{pmatrix}, \quad (58)$$

where $|\varepsilon_{x,y,z,w}| \ll 1$, while for leptons we obtained a near TBM fit (56) with

$$U_{PMNS}^{DK} \approx \begin{pmatrix} 1 & 0 & 0 & 0 \\ 0 & 1/\sqrt{2} & 1/\sqrt{2} & 0 \\ 0 & -1/\sqrt{2} & 1/\sqrt{2} & 0 \\ 0 & 0 & 0 & 1 \end{pmatrix} \begin{pmatrix} \sqrt{2/3} & \sqrt{1/3} & 0 & 0 \\ -\sqrt{1/3} & \sqrt{2/3} & 0 & 0 \\ 0 & 0 & -1 & 0 \\ 0 & 0 & 0 & \pm 1 \end{pmatrix} \begin{pmatrix} 1 & 0 & 0 & 0 \\ 0 & 1 & 0 & 0 \\ 0 & 0 & 1 & 0 \\ 0 & 0 & 0 & 1 \end{pmatrix}. \quad (59)$$

What causes such spectacular difference between quarks and leptons in the middle factor of (58, 59) is not clear. Also puzzling is that the Bogoliubov factors for leptons, although clearly not arbitrary, do not follow the canonical form given by (34). We speculate that this particular puzzle would be resolved if we take into consideration that the mass of the fourth neutrino is so heavy that it is not present in the currently observed three neutrino mixing patterns. Also our assumption that neutrinos are Dirac particles may very well play a significant role in resolution of the puzzle. Inclusion of Majorana scenario in the analysis might produce significant changes.

## 4. Summary

In this Letter we completed the tree level description of the quantum DK-SM theory that we begun in [1] by describing the foundations of the theory and continued in [2] by computing the mass spectrum of the theory. There we also derived the four tree level mass relations that allowed us to etimate masses of the fourth generation from the masses of the first three generations. As a result we were able to predict the normal mass hierarchy for neutrinos, and predict the masses of the 4$^{th}$ charged lepton at approx. $370 \, GeV$ and of the two bottom and top 4$^{th}$ generation quarks, at approx. $100 \, GeV$ and $100 \, TeV$, respectively. The prediction for quark masses should be taken in the sense of the order of magnitude prediction only.

In the present Letter we have derived from the first principles the mass eigenstate mixing parameters of the quark sector of the theory. We were able for the first time to make predictions about the parameters of DK-SM that can be compared with the experimental data, namely that at tree level $V_{ts} = V_{cb}$ and $V_{cs} = V_{tb}$. These are the two simplest of algebraic relations that involve the elements of the CKM-3 in DK-SM. The method that we used to construct the mixing matrix for the quark sector can also be used to fit the existing data with the near-tri-bimaximal representation of the lepton mixing sector. However, to achieve a good fit with the experimental data we had to make a puzzling assumption about the nature of Bogoliubov mixing of the leptons. We conjecture that the deviation from the expected canonical form of the DK-SM mixing is caused by the large masses of the 4$^{th}$ generation of leptons and/or by neutrinos being Majorana.

We would like to stress here, that the three Dirac-Kähler fields, which we used to construct the $SU(2)_L$-invariant lagrangian that contains dimension three mass terms with Dirac fields in right-left asymmetric representations, in fact describe a different form of fermionic matter, the Dirac-Kähler matter. It has different properties then the usual Dirac fermionic matter. Unlike the Dirac fields the Dirac-Kähler fields contain information not only about their Dirac constituents but also about mixing of the constituent particles. Therefore, with this in mind, we believe we are justified in making the claim that the Dirac-Kähler fermionic matter generalizes the notion of Dirac fermionic matter.

The use of DK-SM offers other advantages, in addition to the predictive power of the theory about the parameters of the CKM-3 matrix and the 4$^{th}$ generation masses. For example, the use of DK-SM spinors allows one to write down the unique coupling of matter and gravity. Since Dirac spinors do not couple to metric but to the vierbein one-forms $e^a$, there is no unique way in the theory of gravity to introduce interaction between Dirac matter and gravity. In fact, because of this problem, the exact form of interaction interaction between Dirac matter and gravity is undetermined. In contradistinction to Dirac matter, Dirac-Kähler matter couples directly to metric and, therefore, the coupling of Dirac-Kähler matter to gravity is unique.

The second advantage of the use of Dirac-Kähler matter is that it can be defined on an arbitrary manifold. It is well known, that Dirac spinors can only be defined on manifolds that admit spin structure and, for non-compact manifolds, are parallelizable. Nature provides no indication, however, as to how the vanishing of certain characteristic classes, the second Stiefel-Whitney classes, which is needed to define spin structure, is connected to the Physics of our world.

The third advantage of the use of Dirac-Kähler matter is that it allows one to treat the fermions and the bosons on equal footing conceptually. Although supersymmetry has been known for a long time, Dirac spinor fields, described as sections of a certain bundle, and gauge fields, which are most naturally described as collections of connection one-forms, mathematically seem to live in two different worlds. If one uses Dirac-Kähler spinor description for fermionic matter then both fermionic and bosonic degrees of freedom live in the same world of representations of Lie algebra-valued differential forms. This allows one to have a new look at the supersymmetry as a transformation connecting commuting and anti-commuting differential forms and thus could lead to new realizations of supersymmetry.

Finally, it should be mentioned that the Dirac-Kähler matter naturally introduces a non-vanishing vacuum expectation value of the Hamiltonian. This comes about because to diagonalize the mass matrices we are forced to take the anti-symmetrized normal ordering for the corresponding quantum DK-SM action or Hamiltonian, instead of the conventional normal ordering. While the conventional normal ordering always results in zero expectation value of the Hamiltonian, the vacuum expectation value of the anti-symmetrized normal ordered Hamiltonian is always non-zero and is negative. We speculate that the renormalized value of this constant is the source of the cosmological constant that, in turn, generates the observed negative cosmological dark energy that contributes to the expansion of the universe.

Of course, in the final count, whether or not the DK-SM indeed describes a new form of fermionic matter and can serve as a phenomenologically viable extension of the SM still is an open question. The large predicted difference in masses between $t'$ and $b'$ must be explained, in view of strong EW constraints coming from the $U$, $T$, $S$ oblique parameters. The potentially too low $m_{b'}$ problem must be also resolved. We must stress, however, that the order of magnitude 4$^{th}$ quark mass estimates in Fig. 1 cannot be used to rule DK-SM out. To obtain the 4$^{th}$ quark mass predictions that can be compared with the EW precision data one has to compute the masses of light quarks within DK-SM lattice simulations, derive the known SM EW bound on $|m_{t'} - m_{b'}|$ without the use of Higgs field, and apply mass renormalization group analysis to the tree-level mass relations derived in [2]. In order to do this a consistent perturbation theory based on the DK-SM must be developed. We shall address these issues in the following publications.